# Mid-infrared single photon detector with superconductor Mo$_{80}$Si$_{20}$ nanowire

Qi Chen*, Rui Ge*, Labao Zhang, Feiyan Li, Biao Zhang, Yue Dai, Yue Fei, Xiaohan Wang, Xiaoqing Jia, Qingyuan Zhao, Xuecou Tu, Lin Kang, Jian Chen and Peiheng Wu

Research Institute of Superconductor Electronics, Nanjing University 163 Xianlin Road, Nanjing 210023, China
Corresponding to Lzhang@nju.edu.cn

**Abstract:** A mid-infrared single photon detector (MIR-SNSPD) was reported based on 30 nm-wide superconductor molybdenum silicide nanowires in this work. Saturated quantum efficiencies (QEs) were achieved at the wavelength ranging from 1.55 to 5.07 µm in experiments. At the same time, the intrinsic dark count rate (DCR) was below 100 cps. Thus, this device produced a noise equivalent power (NEP) of $4.5 \times 10^{-19}$ W/sqrt(Hz). The results provide the foundation of developing 10 µm-SNSPD for the applications of infrared astronomy observation.

## 1. Introduction

Mid-infrared single-photon detectors have broad application prospects in many fields, such as biomolecular spectrum analysis [1,2] and astronomical observations [3]. Up to now, there are several photodetectors, including the third-generation semiconductor photodetectors HgCdTe [4,5], QWIPs [6] and type-II Superlattices [7,8]. Unfortunately, it is a big challenge to achieve high efficiency for single-photon detection in the mid-infrared band, resulting in a typical noise equivalent powers (NEPs) on the order of pW/sqrt (Hz). Superconducting nanowire single-photon detectors (SNSPDs) have recently emerged as the highest-performing single-photon detectors due to near-unity efficiency in the near-infrared band [9], high count rate [10,11], dark count rates as low as $10^{-3}$ cps [12,13], and low timing jitter (<3ps) [14].

Recently, the SNSPDs with amorphous WSi showed great potential in mid-infrared applications [15,16], which motivates SNSPDs to develop interesting mid-infrared detection investigations. In 2011, SNSPD achieved 0.5-5 µm single-photon detection based on 30 nm NbN, but the QE was not completely saturated at wavelengths longer than 3 µm [17]. After that, the same authors [18] prepared an SNSPD at 2.1-5.5 µm wavelength range based on the low energy gap material WSi (2Δ(0) ≈1.52 meV). Furthermore, Verma et al. [16] have been reported a 50 nm-wide WSi SNSPD at the wavelength from 2 µm to 7.4 µm. Taylor et al. [19] realized deep lidar imaging at 2.3 µm wavelength based on 60 nm wide NbTiN nanowires, but the detection efficiency is lower than 2%.

However, compared to near-infrared photons, the photon at mid-infrared wavelength has lower energy. Therefore, the quantum efficiency (QE) of SNSPD hugely degenerated in mid-infrared wavelength, limiting the development of MIR-SNSPD. How to effectively improve both the QE and sensitivity has become a crucial topic in single-photon detection technology and scientific applications.

This study demonstrated a high-performing MoSi-SNSPD with a uniform width of 30 nm, which shows a saturated quantum efficiency at the wavelength range from 1.55 μm to 5.07 μm.

## 2. Device fabrication

The nano-fabrication process is similar to our previous work [20]. A double-sided nitrided silicon substrate was adopted, the MoSi superconducting film ($Mo_{80}Si_{20}$ alloy target) was deposited on the upper silicon nitride surface. In order to reduce the oxidation of the MoSi film, the $Nb_5N_6$ capping layer was deposited on MoSi surface. Figure 1(c) shows the cross-section of the MoSi film taken by transmission electron microscopy (TEM), the thickness of the MoSi film and the $Nb_5N_6$ capping layer are 6.08 nm and 3.16 nm, respectively. The square resistance of the MoSi film measured by the four-terminal method is 248.6Ω/□, and the superconducting critical temperature $T_c$ measured under liquid helium is 4.5 K (the temperature corresponding to the resistance at 10% $R_{20K}$).

During the nano-fabrication process, the SNSPD is vulnerable to constrictions with a decrease of the nanowire width. Therefore, it is necessary to optimize the fabrication processes (such as electron beam lithography (EBL), development, and RIE, etc.) to have the narrow MoSi nanowires with excellent performance. Compared with other electron beam resists (such as PMMA2, ZEP520, and AR6200.04, etc.), HSQ has extremely high resolution and excellent etching resistance [21]. Here, we adopted HSQ (solute concentration 2%) as the electron beam resist for writing nanowires through EBPG5200 EBL system. A 100 pA writing beam was selected to improve the accuracy of EBL. Furthermore, we adjusted the electron energy to 100 keV, so that the beam spot size of 100 pA can be reduced to less than 10 nm. Both the writing resolution and BSS were set to 2 nm.

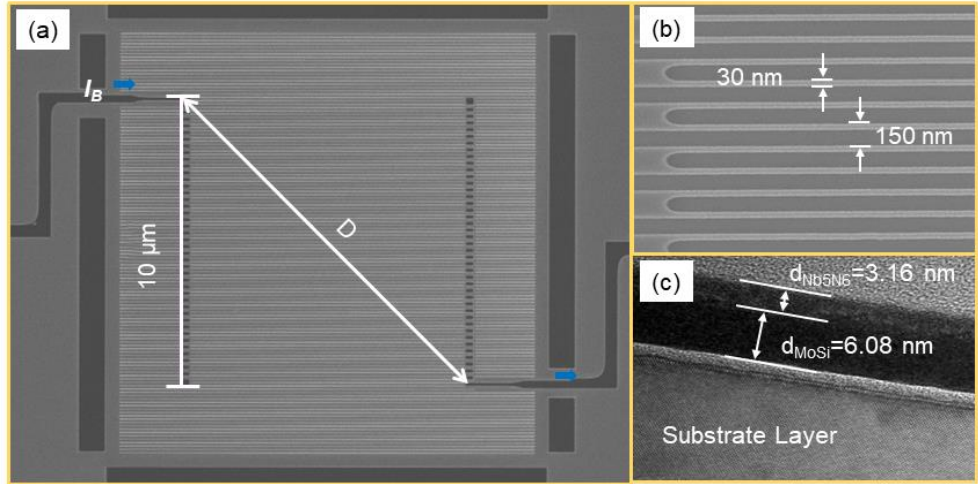

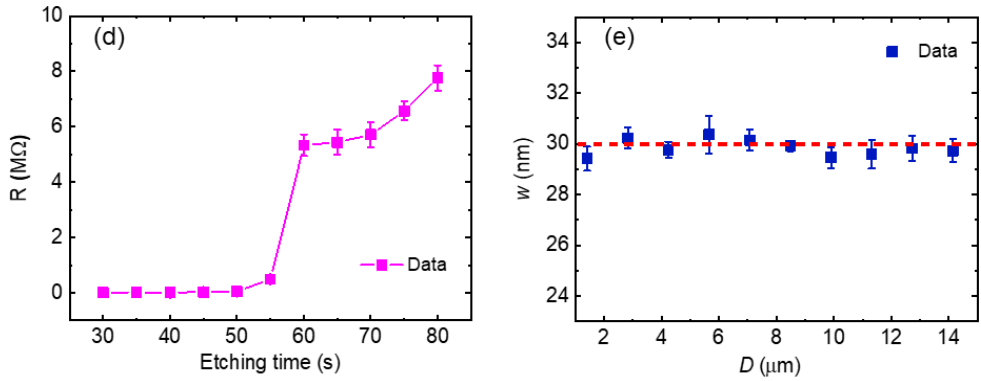

Figure 1. MoSi-SNSPD nano-fabrication. (a) The SEM image of the device active area with 10 μm × 10 μm. The dark blue arrow indicates the direction of the bias current flow. (b). The SEM image of the MoSi nanowire, which enlarged from (a). (c). The TEM image of the cross-section of MoSi film. (d). The relationship between the device's resistance and the etch time during the reactive ion etching (RIE) process. (e). SEM measures the nanowire width fluctuation at different positions on the diagonal line of the active area. The red dash line shows the nanowire width of 30 nm.

On the other hand, appropriately reducing HSQ coating's thickness would reduce the forward scattering of high-energy electrons to finally improve the writing resolution. Considering the accuracy of the writing image and the etching resistance of HSQ, we decided to reduce the HSQ coating's thickness to 30 nm. The pre-baking temperature of HSQ is 90°C, and the baking time is 4 min. After written by EBL, it is developed with 2.38% TMAH developer at room temperature (23°C) for 3 minutes. During the development process, the substrate should be prevented from shaking violently to cause the images to "drift". The nanowire image was transferred to the MoSi film by RIE. The optimized RIE process is as follows: $CF_4$ gas, flow rate 20 sccm, pressure 1.2 Pa, and etch power 50 W. Figure 1(d) shows the relationship between the device's resistance and the etch time during RIE process. It can be seen that the nanowires were just etched suitably when the etching time was 60 s. In order to reduce the edge roughness of the nanowires, we appropriately extended the etching time to 65 s. Figure 1(a) is the active area's SEM image, which shows size of 10 μm × 10 μm. The dark blue arrow indicates the direction of the bias current flow. It can be seen from Figure 1(b) that the actual nanowire width and pitch are 30 nm and 150 nm, respectively, which are consistent with the design values. To characterize MoSi nanowires' uniformity, we first selected the diagonal line of the active area (the length of the diagonal line is D ≈ 14.14 μm). Then we defined any end of the line as the origin. Starting from the origin, we measured the width of the nanowires at different positions on the line. Figure 1(e) shows that the MoSi nanowires in the active area have a good homogeneity.

## 3. Measurements and discussions

Figure 2 shows the experimental schematic of the quantum efficiency (QE) measurement of the MoSi-SNSPD. For optical characterization, we use a silicon nitride blackbody source with an operating temperature of 1500 K, generating a continuous mid-infrared radiation spectrum (Cool-red Ocean Optics). And The maximum output power is 50 W. The optical intensity was controlled with a circular, metallic variable neutral-density filter. The bandpass filters are capable of replacing to ensure that only the specific wavelength can illuminate the SNSPD. In the experiment, we adopted seven kinds of bandpass filters(bought from the Thorlabs company), which have center wavelengths at 1.55 μm, 2.00 μm, 2.25 μm, 3.00 μm, 4.00 μm, 4.26 μm, and 5.07 μm, respectively. These bandpass filters were placed on the optical path in front of the dilution refrigerator's glass window to select a specific wavelength. The light was collimated and free-space coupled into the cryostat through the bandpass filter and the vacuum chamber windows, respectively. Front illuminating the device installed in the dilution refrigerator at a base temperature of 80 mK. The response electrical pulse signal generated by the response nanowire is amplified and finally processed by a time-correlated single-photon counter (TCSPC).

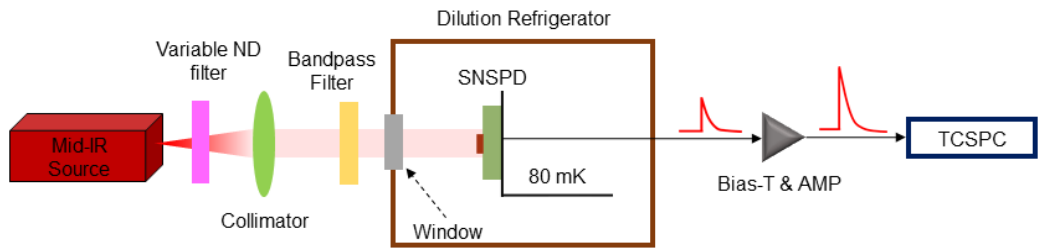

Fig.2 Experimental schematic of quantum efficiency measurement. The blackbody radiation source was used to generate a continuous

midinfrared radiation spectrum. The optical intensity was controlled with a circular, metallic variable neutral-density (ND) filter. The radiation was collimated and then illuminated the bandpass filter. The bandpass filter was capable of replacing to ensure that only the specific wavelength can illuminate the SNSPD. Filters used in the experiment have center wavelengths at 1.55 μm, 2.00 μm, 2.25 μm, 3.00 μm, 4.00 μm, 4.26 μm and 5.07 μm, respectively. The investigation was carried out by using a dilution refrigerator at a base temperature of 80 mK. The signal from the SNSPD was amplified by the amplifier (AM-1309 MITEQ INC) mounted at room temperature. The amplifier has a nominal gain of 50 dB, a bandwidth of 1.5 GHz with a low-frequency cut-off at 1 MHz. The SNSPD was biased with a low-noise current source through a resistive bias-T at the amplifier's input. The amplified signal would be counted in a TCSPC module.

We characterized the electrical properties of MoSi-SNSPD in the dilution refrigerator (80 mK). It has a switching current ($I_{sw}$) of 4.34 μA, When the device is biased at $I_B = 0.9\, I_{sw}$ (where $I_B$ is the bias current of the nanowires). The full width at half maxima (FWHM) of the response pulse signal is 28 ns, and the theoretical maximum count rate of the corresponding device is about 35.7 MHz. Besides, the device's signal-to-noise ratio is about 8.4 times, making it easy to distinguish signals from noise.

We used the InGaAs photodiode power sensor (S148C Thorlabs) to calibrate the wavelength from 1.55 μm to 2.25 μm and used the HgCdTe integrating sphere photodiode power sensor (S180C Thorlabs) to calibrate the wavelength from 3.00 μm to 5.07 μm. To calibrate the light intensity, we place the power sensor at the same device's position in the optical setup. The power sensor receiving port is guaranteed to be on the optical axis. Defined the optical power ratio to the aperture area of the probe as the photon density ($\rho$) in the parallel infrared beam. Adjusting the metallic variable neutral-density (ND) filter to attenuate the optical power to the single-photon level according to the measured value (or calculated photon density). The attenuation rate of the metallic variable neutral-density (ND) filter is calculated according to the measured power and the target single-photon level for different wavelengths.

Combined with the active area, the number of received photons on the area per second $n = \rho \times (10\ \mu m)^2$ could be obtained. $n$ was kept below half of the theoretical maximum count rate of the detector to avoid 'latching' state. At the temperature of 80 mK, we measured the quantum efficiencies (QEs) of different infrared photons with wavelengths of 1.55 μm, 2.00 μm, 2.25 μm, 3.00 μm, 4.00 μm, 4.26 μm, and 5.07 μm, respectively, as shown in Figure 3. QE represents the probability that a nanowire absorbs one photon and then to produce a detectable electrical signal. In this work, we define $QE = PCR(I_B) / PCR_{saturation}$, where $PCR(I_B)$ is the photon count rate related to the bias current $I_B$, and $PCR_{saturation}$ represents the saturated count rate that is independent of $I_B$. It was demonstrated that there is an "S"-shaped relationship between the detection efficiency and the bias current [22]. Therefore, we use the "S" equation to fit the relationship between QEs and $I_B$ at different wavelengths, the equation as follows(1):

$$QE(I_B) = \frac{1}{1+e^{-a(I_B-b)}} = \frac{e^{a(I_B-b)}}{e^{a(I_B-b)}+1} \quad (1)$$

Both *a* and *b* are the fitting parameters. The solid lines of different colors represent the fitting curves at different wavelengths. It can be seen that the QEs can perfectly reach 100% in the bias range of $I_B < I_{SW}$ at the wavelength of λ < 4.26 μm. Notably, the QEs still tend to be saturated when λ > 4.26 μm. The maximum QE can reach 98% at the wavelength of λ = 4.26 μm and can reach up to 97% at the wavelength of λ = 5.07 μm. According to the fitting curves, it can be inferred that if the bias current $I_B$ > 4.6 μA, saturated QEs could be achieved at the wavelengths from 1.55 to 5.07 μm. Therefore, for the purpose of eliminating "constrictions" [23] and further increasing the $I_{SW}$, it is essential to continue to optimize the design and nano-fabrication to improve the homogeneity of nanowires.

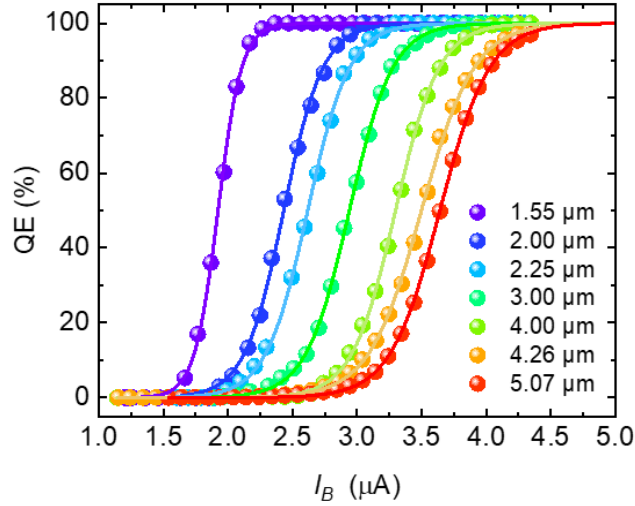

Figure 3. The relationship between quantum efficiencies (QEs) and bias current ($I_B$) at the base temperature of 80 mK. Different colors represent different working wavelengths, the solid circles are experimental measurement data, and the solid curves represent fitting results of the "S" equation to QEs for different wavelengths. The superconducting critical current ($I_{SW}$) is 4.34 μA.

Furthermore, we analyzed the noise equivalent power (NEP) of the device. Generally, NEP is defined as the amount of infrared radiation power that makes the signal-to-noise ratio equal to 1. When the power is less than NEP, the detector cannot sense the target radiation. Therefore, NEP is actually the smallest target radiation that the detector can detect, and it marks the sensitivity of a detector. The smaller the NEP, the higher the sensitivity. For single-photon detectors, the calculation of NEP is as the following equation (2):

$$\text{NEP} = \frac{\sqrt{2}E_\lambda}{QE} \times \sqrt{DCR} \qquad (2)$$

Where $E_\lambda$ represents photon energy. In this experiment, we measured the intrinsic dark count rate (DCR) under the environmental electromagnetic shielding of the detection system, the DCR of the device is no more than 100 cps, as shown in Figure 4(a). The optimal NEPs of the device at different wavelengths were evaluated according to equation (2), as described in Figure 4(b). It can be seen that these NEPs corresponding to the wavelengths (1.55-5.07 μm) are always less than $4.5 \times 10^{-19}$ W/sqrt(Hz), confirming the extremely high sensitivity of the device.

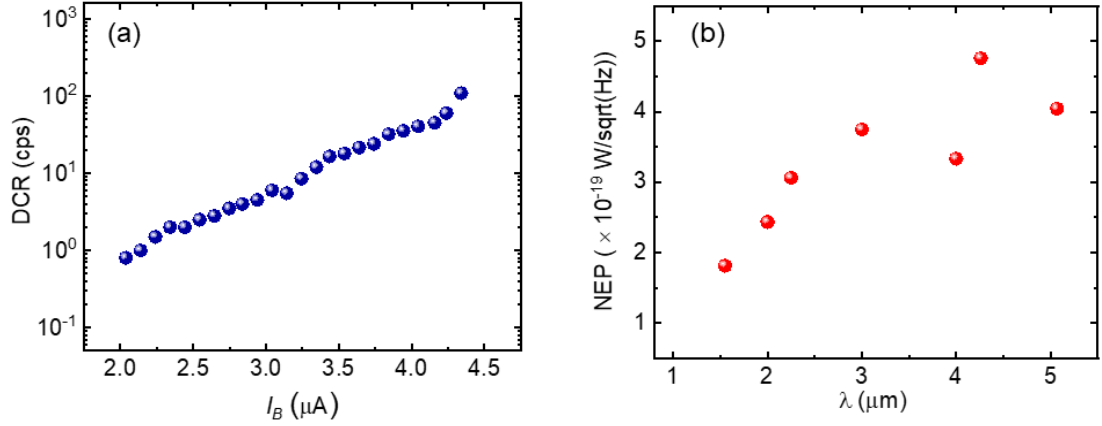

Figure 4. (a). The measured dark count (DCR) vs $I_B$ under the environmental electromagnetic shielding of the detection system. The maximum DCR is about 100 cps. (b). Noise equivalent powers (NEPs) of the device at different wavelengths, which are always lower than $4.5 \times 10^{-19}$ W/sqrt(Hz) at the 1.55-5.07 μm wavelength range.

## 4. Conclusions

In summary, we reported a MoSi-SNSPD with saturated quantum efficiency at 1.55-5.07 μm wavelength range. The experimental results show that the QEs can reach 100% in the bias range of $I_B < I_{SW}$ at the wavelengths of λ < 4.26 μm. When λ > 4.26 μm, QEs still tend to be saturated (98%@4.26 μm and 97%@5.07 μm). The noise equivalent powers (NEPs) are always lower than $4.5 \times 10^{-19}$ W/sqrt(Hz) at 1.55-5.07 μm wavelength range. The results of this study have prompted the development of physics research of MIR-SNSPD. They also have provided potential applications for the development of infrared astronomical detection and other scientific fields.

**Acknowledgments:** This work was supported by the National Key R&D Program of China Grant (2017YFA0304002), the National Natural Science Foundation (Nos. 12033002, 61571217, 61521001, 61801206 and 11227904), the Priority Academic Program Development of Jiangsu Higher Education Institutions (PAPD), and the Jiangsu Provincial Key Laboratory of Advanced Manipulating Technique of Electromagnetic Waves.

# Reference

1. Jascha A. Lau, Arnab Choudhury, Li Chen et al. Observation of an isomerizing double-well quantum system in the condensed phase. Science 367, 175-178 (2020).

2. L. Chen, D. Schwarzer, V. B. Verma et al. Mid-infrared Laser-Induced Fluorescence with Nanosecond Time Resolution Using a Superconducting Nanowire Single-Photon Detector : New Technology for Molecular Science. Acc. Chem. Res. 50, 1400-1409 (2017).

3. M. E. R. Essler, K. G. S. Ukhatme, B. R. F. Ranklin et al. The Mid-Infrared Instrument for the James Webb Space Telescope, VIII : The MIRI Focal Plane System. Soc. Pacific. 127, 675–685 (2015).

4. Qiu Weicheng, Hu Weida, Lin Chun et al. Surface leakage current in 12.5 mu m long-wavelength HgCdTe infrared photodiode arrays. Opt. Lett. 41, 828-831 (2016).

5. J. B. Abshire, M. A. Krainak, W. Lu et al. HgCdTe avalanche photodiode array detectors with single photon sensitivity and integrated detector cooler assemblies for space lidar applications. Opt. Eng. 58, 067103 (2019).

6. A. Rogalski. Quantum well photoconductors in infrared detector technology. Journal of Applied Physics 93, 4355 (2003).

7. A. Rogalski, P. Martyniuk, and M. Kopytko. InAs / GaSb type-II superlattice infrared detectors : Future prospect. Appl. Phy. Rev. 4, 031304 (2017).

8. I. Vurgaftman, E. H. Aifer, C. L. Canedy et al. Graded band gap for dark-current suppression in long-wave infrared W- structured type-II superlattice photodiodes. Appl. Phys. Lett. 89, 121114 (2006).

9. F. Marsili, V. B. Verma, J. A. Stern et al. Detecting single infrared photons with 93 % system efficiency. Nat. Photonics **7**, 210-214 (2013).

10. W. H. P. Pernice, C. Schuck, O. Minaeva et al. High-speed and high-efficiency travelling wave single-photon detectors embedded in nanophotonic circuits. Nat Communications 3, 1325 (2012).

11. J. Huang, W. Zhang, L. You et al. High speed superconducting nanowire single-photon detector with nine interleaved nanowires. Supercond. Sci. & Tech. 31, 074001 (2018).

12. H. Shibata, K. Shimizu, H. Takesue et al. Ultimate low system dark-count rate for superconducting nanowire single-photon detector. Opt. Lett. **40**, 3428–3431 (2015).

13. H. Shibata, K. Fukao, N. Kirigane et al. SNSPD With Ultimate Low System Dark Count Rate Using Various Cold Filters. IEEE Trans. on Appl. Supercond. **27**, 1-4 (2017).

14. B. Korzh, Qing-Yuan Zhao, J. P. Allmaras et al. Demonstrating sub-3 ps temporal resolution in a superconducting nanowire single-photon detector. Nat. photonics 14, 250-255 (2020).

15. Li Chen, Jascha A. Lau, Dirk Schwarzer et al. The Sommerfeld ground-wave limit for a molecule adsorbed at a surface. Science 353, 158-161 (2019).

16. V. B. Verma, A. E. Lita, B. Korzh et al. Towards single-photon spectroscopy in the mid-infrared using superconducting nanowire single-photon detectors. Proc. SPIE 10978, Advanced Photon Counting Techniques XIII, 109780N (2019).

17. F. Marsili, F. Bellei, F. Naja et al. Efficient Single Photon Detection from 500 nm to 5 μm Wavelength. Nano. Lett. 12, 4799-4804 (2012).

18. F. Marsili, V. B. Verma, M. J. Stevens et al. Mid-Infrared Single-Photon Detection with Tungsten Silicide Superconducting Nanowires. CLEO: Science and Innovations. 978-1-55752-972-5 (2013).

19. G. G. Taylor, D. Morozov, N. R. Gemmell et al. Photon counting LIDAR at 2.3 µm wavelength with superconducting nanowires. Opt. Exp. 27, 38147-38158 (2019).

20. Qi Chen, Biao Zhang, Labao Zhang et al. Sixteen-Pixel NbN Nanowire Single Photon Detector Coupled With 300-μm Fiber. IEEE Photon. J. 12, 6800112 (2020).

21. A. E. Grigorescu and C. W. Hagen. Resists for sub-20-nm electron beam lithography with a focus on HSQ: state of the art. Nanotechnology. 20, 292001 (2009).

22. T. Yamashita, S. Miki, H. Terai et al. Low-filling-factor superconducting single photon detector with high system detection efficiency. Opt. Exp. 21, 27177–27184 (2013).

23. Andrew J. Kerman, Eric A. Dauler, Joel K. W. Yang et al. Constriction-limited detection efficiency of superconducting nanowire single-photon detectors. Appl. Phys. Lett. 90, 101110 (2007).